# Laser heating control with polarized light in isolated multi-walled carbon nanotubes

*Mariusz Zdrojek\*, Jarosław Judek, Michał Wąsik*

Faculty of Physics, Warsaw University of Technology, Koszykowa 75, 00-662 Warsaw, Poland

Abstract

We are proposing a novel method of laser heating control only through change in polarization of the incident light, keeping its power density constant. The idea combines antenna effect found in isolated multi-walled carbon nanotubes and the possibility of their heating by light illumination. To observe this we used Raman spectroscopy technique, where the heating manifests itself in a pronounced downshift of the Raman G and 2D lines as a function of the polarization angle. Our method can be useful in field electron emission devices or in selective nanotubes heating and destruction. It can also be extended to other one dimensional nanoobjects, if only certain conditions are fulfilled.

PACS numbers: 61.48.De, 78.20.Ci, 84.40.Ba, 61.80.Ba

\* zdrojek@if.pw.edu.pl

In infrared and visible regime, carbon nanotubes can act as optical antennae. Particularly, they can exhibit the polarization antenna effect that suppresses their optical response when the electric field of the incoming radiation is polarized perpendicular to nanotube axis. Usually, response stands for the elastic [1,2] or inelastic [3-5] light scattering; however, it can be generalized to other effects caused by the light illumination, as we demonstrate in this paper.

Our experiment starts from the identification of the antenna effect in isolated multi-walled carbon nanotubes (MWCNTs) using polarized Raman spectroscopy. We do this by analyzing the G and 2D Raman feature intensities as a function of the angle $\varphi$ between the nanotube axis and the polarization of incident light. Comparison to bundles [6,7] and a reference to the resonance conditions [8] suggest that signal intensity suppression can originate in a depolarization effect[9,10]. Further analysis of the G and 2D modes shows that neither their shapes nor positions are affected by the change in the $\varphi$ angle for low laser power densities (LPDs). For higher LPD, mode positions begin to shift toward lower frequencies because laser light absorption induces thermal expansion of the $sp^2$ lattice [11-16]. However, these shifts for fixed power densities are $\varphi$ dependent! This means that we are able to observe a rise in the local temperature that depends only on the polarization of the incident light due to the anisotropic optical absorption. To our knowledge, this is the first experimental demonstration of polarization dependent heating effect in carbon nanotubes probed by Raman spectroscopy or by any other technique. Potential application includes the control of a nanotube work function in a field electron emission devices by change in the local temperature [17]. By tuning the laser power density, it is also possible to destroy MWCNTs ordered in selected direction keeping the remaining tubes unaffected.

Carbon nanotubes with a diameter $d$ = 15-30 nm were grown using CVD technique and dispersed on the Si/SiO$_2$ substrate. In our experiment, we used Dilor XY-800 Raman spectrometer equipped with a microscope in order to acquire a spectra from individual carbon nanotube (see Figure 1a). Polarizer, analyzer and $\lambda$/2 retardation plate were used to control the direction of the electric field vector of the incoming and outgoing light. All experiments were performed in the backscattering geometry under ambient conditions. For the excitation, 514 nm line from Ar$^+$ laser was used. Calibrated laser power on the sample was changed from 0.7 mW up to 20 mW. In order to verify any variation of laser power on sample as a function of light polarization silicon oriented substrates (100 and 111) for polarization calibration have been used. The laser spot focused with the 100 x objective had a diameter of ~ 2 μm, thus power density varied from ~ 20 kW/cm$^2$ to ~ 600 kW/cm$^2$. Typical MWCNT Raman spectrum is presented in Figure 1b, whereas typical AFM image in Figure 1c.

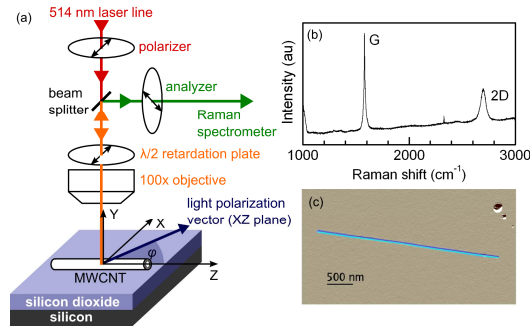

**FIG. 1:** (a) Schematic of the experimental setup used for exploring the dependence of inelastic scattering amplitude and phonon energy on the angle $\varphi$ between carbon nanotube axis and the direction of the electric field vector of the incident $e_i$ and scattered $e_s$ light. For VV configuration $e_i \| e_s$, for VH configuration $e_i \perp e_s$. (b) Raman spectrum from an isolated multi-walled nanotube. (c) AFM image of isolated MWCNT ($d_{isolated}$ = 30 nm) on the SiO$_2$/Si substrate.

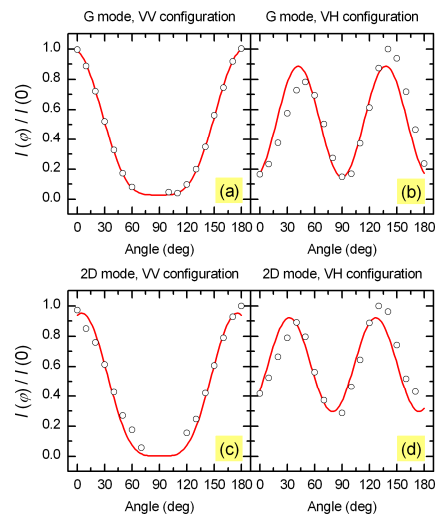

**FIG. 2:** Normalized integrated intensity for VV and VH configuration for G (a, b) and 2D (c, d) bands. Experimental data (open symbols) follow theoretical predictions (lines). Unequal maxima and weaker extinction of intensity for VH configuration are attributed to the light polarization misalignment [*].

Figure 2 shows the G and 2D band intensity dependencies on the angle $\varphi$ between the carbon nanotube axis and the direction of the electric field vector of the incident $e_i$ and scattered $e_s$ light for VV ($e_i \| e_s$) and VH ($e_i \perp e_s$) configuration. Experimental data in Figure 2a and 2c (VV configuration) are well described by $\cos^4\varphi$ function. Data in Figure 2b and 2d (VH configuration) are described by $\sin^2 2\varphi$. Band intensities on all four graphs reach its minimum when the incident or scattered light polarization vector is perpendicular to the nanotube axis ($\varphi_{min} \sim 90°$ for both configurations, for VH also $\varphi_{min} \sim 0°$). It means that inelastic photon scattering occurs mainly for the light polarized along the tube axis, exactly as in a dipolar antenna. The detailed comparison between our results for the isolated multi-walled nanotube and bundles of MWCNTs [6] reveals two significant differences. First, the Raman bands are described only by one Lorentzian function and the changes in their shape are never observed. It means that the inner diameter of our tube is too large for splitting of the G mode. Second, the minimum of the G band intensity has shifted from $\varphi_{min} \sim 55°$ for bundles to $\varphi_{min} \sim 90°$ in our case. This shift can be explained by selective suppression of Raman scattering. To show this, we express the line intensity $I$ using the concept of the Raman tensor $\mathbf{R}$

$$I \propto \left| \mathbf{e}_i \cdot \mathbf{R} \cdot \mathbf{e}_s \right|^2. \tag{1}$$

Literature[6,7] provides the following form of the Raman tensor for a G mode

$$\mathbf{R} \propto \begin{pmatrix} -1/2 & 0 & 0 \\ 0 & -1/2 & 0 \\ 0 & 0 & 1 \end{pmatrix}. \tag{2}$$

Such tensor well describes the experimental data for the MWNCT bundles [6], despite it is derived from non-resonant Raman scattering [7]. Now, we introduce the positive attenuation ratio $c < 1$ for light polarized perpendicular to the nanotube axis that implies

$$\mathbf{R} \propto \begin{pmatrix} -c/2 & 0 & 0 \\ 0 & -c/2 & 0 \\ 0 & 0 & 1 \end{pmatrix}. \tag{3}$$

Next, we can write the equation for a normalized band intensity for the VV configuration

$$\frac{I^{VV}(\varphi)}{I^{VV}(0°)} = \left(\cos^2(\varphi) - \frac{c}{2}\sin^2(\varphi)\right)^2. \qquad (4)$$

When there is no attenuation, $I^{VV}(90°)/I^{VV}(0°) = 1/4$ and $\varphi_{min}$ = arc cos $(1/\sqrt{3})$ ~ 55° [6]. In the case of strong attenuation, $I^{VV}(90°)/I^{VV}(0°) \to 0$ and $\varphi_{min} \to 90°$. Experimental data show that we are in the second regime. Cautious estimation of the attenuation ratio $c$ yields values below 0.25 and $I^{VV}(90°)/I^{VV}(0°)$ is below 0.04. The exact value of $c$ is difficult to obtain due to the very low signal intensity for $\varphi$ ~ 90° and not precise enough polarization control.

Raman scattering from multi-walled carbon nanotubes in the visible range is resonant; therefore, we attribute attenuation of the Raman scattering in the perpendicular direction to the suppression of the absorption. Discrepancies between our results and those obtained for the bundles can be explained by the depolarization effect due to stronger localization of electronic states in the perpendicular direction ($d_{isolated}$ ~ 30 nm << $d_{bundle}$ ~ 1μm)[10] and staying within the electrostatic limit [18]. According to Marinopoulos [10], the suppression of an optical absorption occurs only in the completely isolated nanoobjects because when the tubes are interacting with each other, the electronic states start to delocalize, depolarization weakens and the system is similar to a bulk metal. Another possible explanation was presented by Zhang [18], who found that strong anisotropy due to the depolarization observed for nanowires disappears when their diameters approach the value of the incident light wavelength. In our case, the electrostatic limit was not reached ($d_{isolated}$ ~ 30nm << $\lambda$ ~ 500 nm), whereas, for instance, the limit for multi-walled bundle was greatly exceeded ($d_{bundle}$ ~ 1000 nm >> $\lambda$ ~ 500 nm). The polarization effect was also observed in the single-walled nanotubes [3-5]. However, here, the antenna behavior seems to be an effect of the specific resonance conditions that are different for parallel and perpendicular direction and change with the photon energy [8]. Generally, in SWCNTs the intensity ratio, $I^{VV}(90°)/I^{VV}(0°)$, can assume values larger or smaller than 1, depending on the type of resonance. The disagreement with the expected depolarization effect is explained as a characteristic resonance feature in the single-walled nanotubes. In our experiment, there are no specific resonance conditions because the electronic structure of nanotube with such large diameter resembles the electronic structure of a rolled graphite.

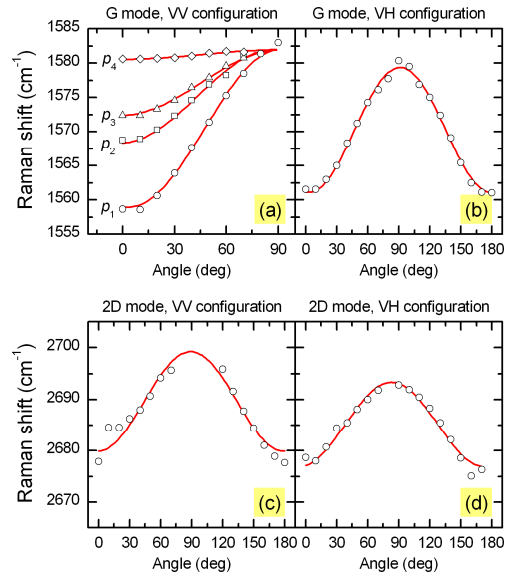

**Figure 3.** Angular evolution of G and 2D band positions for both polarization configurations. Four data series in a plot (a) acquired for the different laser power densities ($p_1 > p_2 > p_3 > p_4$) prove the thermal origin of the Raman shift change. Experimental data (open symbols) were fitted with the $\cos^2(\varphi)$ function (lines).

Angular dependencies of the G and 2D band positions for both VV and VH configuration are presented in Figure 3. The Raman shift $\hbar\omega$ depends on the $\varphi$ angle between the electric field vector and the nanotube axis and is well described by the cosine squared function:

$$\hbar\omega(\varphi) = \hbar\omega(90°) - A \cdot \cos^2(\varphi) \tag{5}$$

For the G band $\hbar\omega(90°) = 1582$ cm$^{-1}$, for the 2D band $\hbar\omega(90°) = 2700$ cm$^{-1}$, $A$ stands for maximal change in Raman shift obtained for $\varphi \sim 0°$ and is different for the G and 2D band. Figure 3a depicts the influence of a laser power density $p$ on value of $A$. Higher LPD implies larger changes in peak positions. For the lowest power $p_4$, only slight peak shifts are observed. Moreover, the shapes of the G and 2D bands can be described by single Lorentzian functions showing no effects on the band symmetry and only their centers are $\varphi$ dependent. Therefore, it is neither a symmetry-related evolution nor an artifact. In Figures 3b-3d only results for one laser power density were shown for sake of clarity. We note that in the existing literature on the polarization dependence of the Raman lines for carbon nanotubes there is no indication of frequency shifts.

Interpretation of the observed phenomena is based on the fact that the position of the G band depends on the local sample temperature [11-16]. Local temperature depends on the efficiency of the laser heating that is the function of absorption and thermal conductivity. Because the latter is constant (we illuminate the same nanotube all the time), the experimentally obtained dependence of the G band position reflects the relation between the MWCNT absorption rate and the polarization angle. This means that we harness the strong anisotropy of optical absorption to control local the sample temperature which is observed as the angular evolution of the position of G and 2D modes. The results obtained in a VH configuration resemble those for the VV because heating by the light absorption depends only on the polarization of the incident light and not on the scattered.

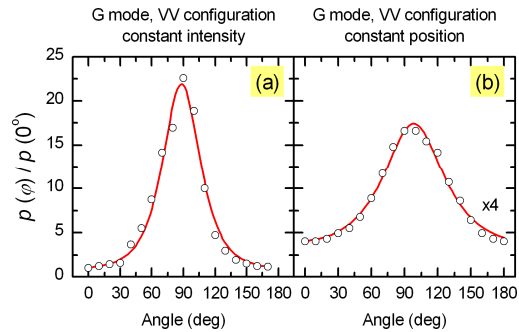

**Fig. 4:** Normalized laser power density required to keep the G mode intensity (a) and position (b) constant. Experimental points (open symbols) are assisted by theoretical curves (lines). In the right picture mode position 1570 cm$^{-1}$ was set.

We now turn our attention to the evaluation of the optical anisotropy by conducting two experiments. In the first one, $\varphi$ angle was changed gradually and at each step the laser power density was increased in order to keep the G mode intensity constant (Figure 4a). Thus normalized laser power density $p(\varphi)/p(0°)$ can be expressed as an inverse equation (4). Maximal power (reaching 600 kW/cm$^2$) was used for $\varphi \sim$ 90°, and we got $p(90°)/p(0°) = 22.6$. This value corresponds with the data shown in Fig. 2a. In the second experiment, light polarization was set along the nanotube axis and the laser power density was adjusted to value $p(0°)$ for which the G mode position was 1570 cm$^{-1}$. Afterwards, $\varphi$ was increased and at each point and we compensated $p$ to keep the G mode position constant (Figure 4b). Normalized LPD in this case should follow:

$$\frac{p(\varphi)}{p(0°)} = \left(\cos^2(\varphi) + c_{pos} \cdot \sin^2(\varphi)\right)^{-1}, \qquad (6)$$

where $c_{pos}$ stands for attenuation of absorption for the direction perpendicular to the nanotube axis. Fit of equation (6) to experimental points (Figure 4b) yields $c_{pos} = 0.23$. Whereas value of $p(\varphi)/p(0°)$ ratio, while keeping the G mode intensity constant (Figure 4a), is equivalent to the typical polarized Raman study presented in Figure 2, the evaluation of the LPD increase, while keeping the G mode position constant (Figure 4b), allows us to observe directly the anisotropy of optical absorption. In details, Raman scattering is a third order process including the photon absorption and emission. In order to keep the line intensity constant, the laser power should be adjusted due to change both in the photon absorption and emission. To keep the G mode position constant, LPD had to be adapted only to decrease in absorption coefficient. In addition, it should be noted that $c_{pos}$ coefficient can be underestimated because for such high laser power densities the vicinity of MWCNT can also be heated thus lowering heat outflow.

Next, we estimated the local temperature arising in the nanotube upon polarized light illumination. According to the literature data for multi-walled nanotubes [11], temperature coefficient for G mode is $\chi_G = -0.028$ cm$^{-1}$/K. In our experiment, the maximal change in Raman shift equaled 24 cm$^{-1}$. This corresponds to 860°C temperature rise in the material. Such temperatures should be treated with care because the tubes have different $\chi$ depending on the growth technique and amount of defects. Despite of that, we observed that several nanotubes were damaged for LPD above 200 kW/cm$^2$. For the tubes that have been damaged we observed substantial decrease in the Raman bands intensity (G, 2D) and appearance of the defect induced D peak.

In summary, we observed the antenna effect in the polarized Raman spectra on the isolated multi-walled nanotubes. The discrepancies between the isolated and bundled MWCNTs are attributed to the depolarization effect. Our study reveals that the change in the position of the Raman modes upon the rotation of the light polarization is not symmetry but heating-related. We ascribe this behavior to the optical absorption anisotropy. We expect that the effect presented here can be found in other high aspect ratio nano-objects, if only localization of the electronic states is high enough[10] or/and they stay within the electrostatic limit [18].

ACKNOWLEDGMENT

The authors acknowledge support from the Foundation for Polish Science (Homing Plus). The authors also thank A. Bachtold and M.J. Esplandiu (Barcelona) for the help in the fabrication of the samples, and L. Wirtz and R. Bacewicz for a critical reading of the manuscript.

[*] In real optical setups direction of the electric field vector of laser light can be diffused around some value. Therefore, the function describing angular evolution of the Raman feature intensities should be convoluted with, *e.g.*, the Gaussian distribution. Particularly, data in Figure 2b and 2d do not have zero values for φ~0° and ~90° due to the polarization misalignment.